\documentclass{article}

\usepackage{arxiv}

\usepackage[utf8]{inputenc} 
\usepackage[T1]{fontenc}    
\usepackage{hyperref}       
\usepackage{url}            
\usepackage{booktabs}       
\usepackage{amsfonts}       
\usepackage{nicefrac}       
\usepackage{microtype}      
\usepackage{lipsum}		    
\usepackage{graphicx}
\usepackage[numbers,square]{natbib}
\usepackage{doi}
\usepackage{amsmath}
\usepackage[bottom]{footmisc}
\usepackage{footnote}
\hypersetup{hidelinks}

\title{Between Generating Noise and Generating Images: Noise in the Correct Frequency Improves \\the Quality of Synthetic Histopathology Images \\for Digital Pathology}

\date{}


\author{\href{https://orcid.org/0000-0002-0939-3379}{\includegraphics[scale=0.06]{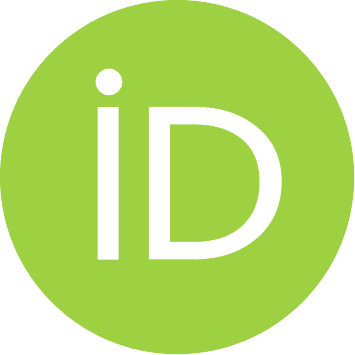}\hspace{1mm}Nati Daniel$^{1}$} \\
    Technion - IIT\\
		\And
  \href{https://orcid.org/0000-0003-4591-7018}{\includegraphics[scale=0.06]{orcid.pdf}\hspace{1mm}Eliel Aknin$^{1,2}$} \\
	Technion - IIT\\
		\And
  \href{https://orcid.org/0000-0002-5006-9300}{\includegraphics[scale=0.06]{orcid.pdf}\hspace{1mm}Ariel Larey$^{1,3}$}\\
    Technion IIT\\
	    \And
     \href{https://orcid.org/0000-0002-2506-3249}{\includegraphics[scale=0.06]{orcid.pdf}\hspace{1mm}Yoni Peretz$^{4}$} \\
	Technion - IIT\\
		\And
  \href{https://orcid.org/0000-0003-4339-6167}{\includegraphics[scale=0.06]{orcid.pdf}\hspace{1mm}Guy Sela$^{4}$} \\
	Technion - IIT\\
		\And
	\href{https://orcid.org/0000-0003-3304-4854}{\includegraphics[scale=0.06]{orcid.pdf}\hspace{1mm}Yael Fisher$^{5}$} \\
	Rambam Health Care Campus\\
		\And
	\href{https://orcid.org/0000-0002-5345-8491}{\includegraphics[scale=0.06]{orcid.pdf}\hspace{1mm}Yonatan Savir$^{1,}$\thanks{Corresponding author, e-mail: yoni.savir@technion.ac.il. $^{1}$Department of Physiology, Biophysics and System Biology, Faculty of Medicine, Technion Israel Institute of Technology, Haifa, Israel.
   $^{2}$Faculty of Industrial Engineering, Technion Israel Institute of Technology, Haifa, Israel.
    $^{3}$Faculty of Computer Science, Technion Israel Institute of Technology, Haifa, Israel. 
    $^{4}$Faculty of Electrical Engineering, Technion Israel Institute of Technology, Haifa, Israel. 
    $^{5}$Division of Pathology, Rambam Health Care Campus, Haifa, Israel.}} \\
	Technion - IIT\\
}

\date{}

\renewcommand{\headeright}{Daniel N et al.}
\renewcommand{\shorttitle}{Between Generating Noise and Generating Images}

\begin{document}

\maketitle

\begin{abstract}
Artificial intelligence and machine learning techniques have the promise to revolutionize the field of digital pathology. However, these models demand considerable amounts of data, while the availability of unbiased training data is limited. Synthetic images can augment existing datasets, to improve and validate AI algorithms. Yet, controlling the exact distribution of cellular features within them is still challenging. One of the solutions is harnessing conditional generative adversarial networks that take a semantic mask as an input rather than a random noise. Unlike other domains, outlining the exact cellular structure of tissues is hard, and most of the input masks depict regions of cell types. However, using polygon-based masks introduce inherent artifacts within the synthetic images – due to the mismatch between the polygon size and the single-cell size. In this work, we show that introducing random single-pixel noise with the appropriate spatial frequency into a polygon semantic mask can dramatically improve the quality of the synthetic images. We used our platform to generate synthetic images of immunohistochemistry-treated lung biopsies. We test the quality of the images using a three-fold validation procedure. First, we show that adding the appropriate noise frequency yields 87\% of the similarity metrics improvement that is obtained by adding the actual single-cell features. Second, we show that the synthetic images pass the Turing test. Finally, we show that adding these synthetic images to the train set improves AI performance in terms of PD-L1 semantic segmentation performances. Our work suggests a simple and powerful approach for generating synthetic data on demand to unbias limited datasets to improve the algorithms' accuracy and validate their robustness.
\end{abstract}


\section{INTRODUCTION}
Synthetic images of tissues have great potential in facilitating Artificial Intelligence (AI) and machine learning for computational pathology and biomedical applications in general. The ability to control the distribution of scenarios and, by that, debiasing the dataset, allows a better diversity and representation of the training set and validation set \citep{serag2019translational_46, tizhoosh2018artificial_47}. Hence, it enables the development of more accurate and reliable AI models to extract useful information for better diagnoses, clinical outcomes, and treatment decisions, especially in rare disease conditions.

\begin{figure}[thpb]
\centering
	\includegraphics[scale=0.4]{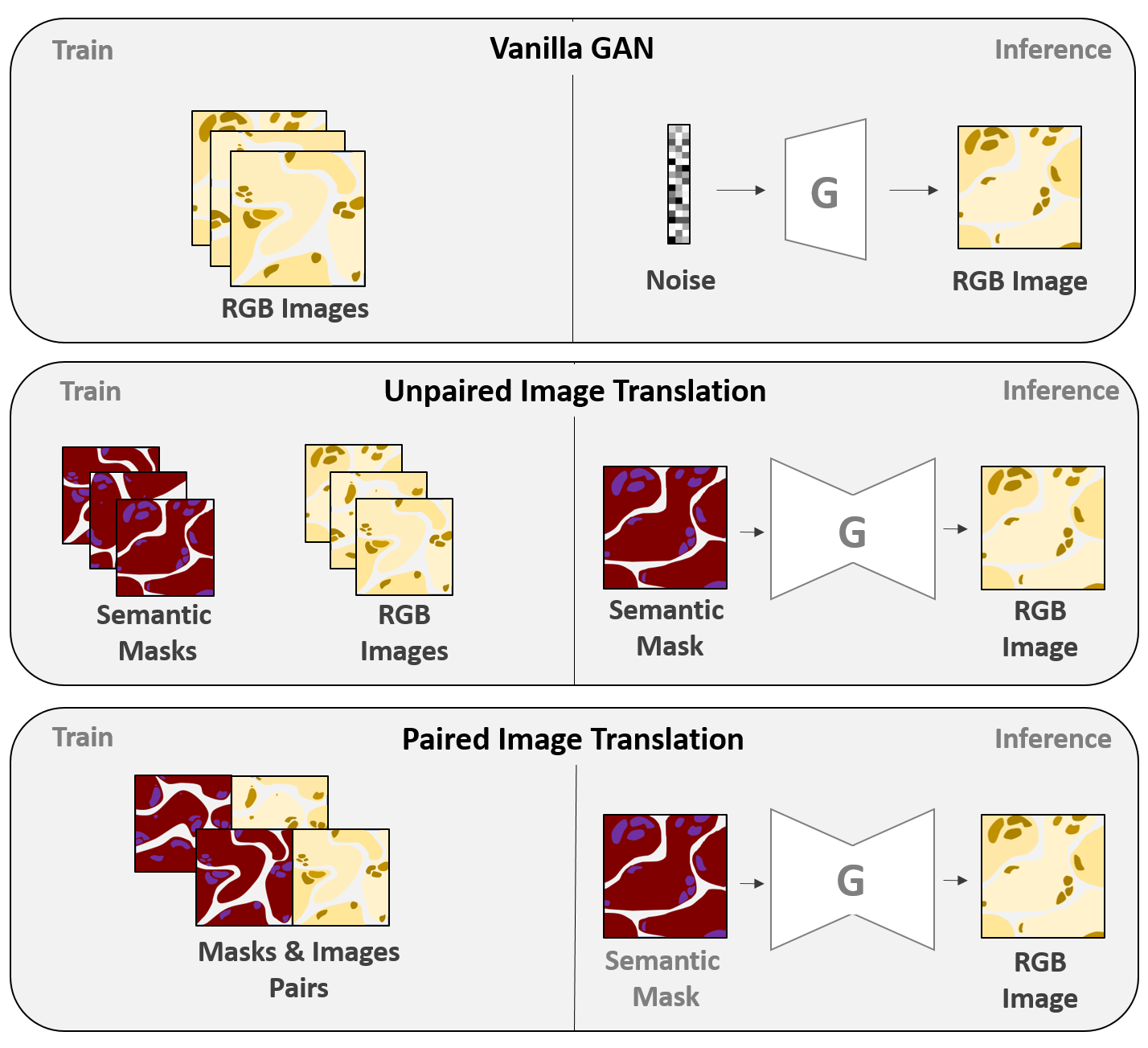}
	\caption{Illustration of three types of generative approaches for producing synthetic images. (Top) Vanilla GAN, random noise is processed by the generator to yield synthetic histology images. On the other hand, image translation approaches (middle and bottom) use prior knowledge by converting discrete semantic labels into RGB photorealistic histology images. The unpaired image translation approach (middle), requires a batch of real images and a batch of semantic masks for the training procedure. While the paired image translation approach (bottom), constrains the images and semantic masks to be paired where each semantic mask is extracted from its corresponding histology image.}
	\label{f:f1}
\end{figure}

Currently, there are three primary approaches exist for generating synthetic images. The first is the classic approach using a generative adversarial network, coined Vanilla GAN, \citep{goodfellow2020generative} in which random noise is processed by the generator to yield synthetic histology images without any prior on the generated images. This approach requires real histology images to be processed by the discriminator during training. The second and third are related to images translation approaches, also known as unpaired and paired image-to-image translation approaches, using conditional generative adversarial network (CGAN) \citep{mirza2014conditional} in which they use prior knowledge by converting discrete semantic label map or other properties into RGB photo-realistic histology images. These approaches also require semantic label maps or other image information during training in addition to the real histology images. Fig.~\ref{f:f1} illustrates both image translation approaches and the classic image generative one. These  approaches result in quality-scalability tradeoffs. The unpaired approach allows for generating a large number of images but lacks the ability to control the cellular features of the image in a precise manner.

\begin{figure}[thpb]
\centering
	\includegraphics[scale=0.4]{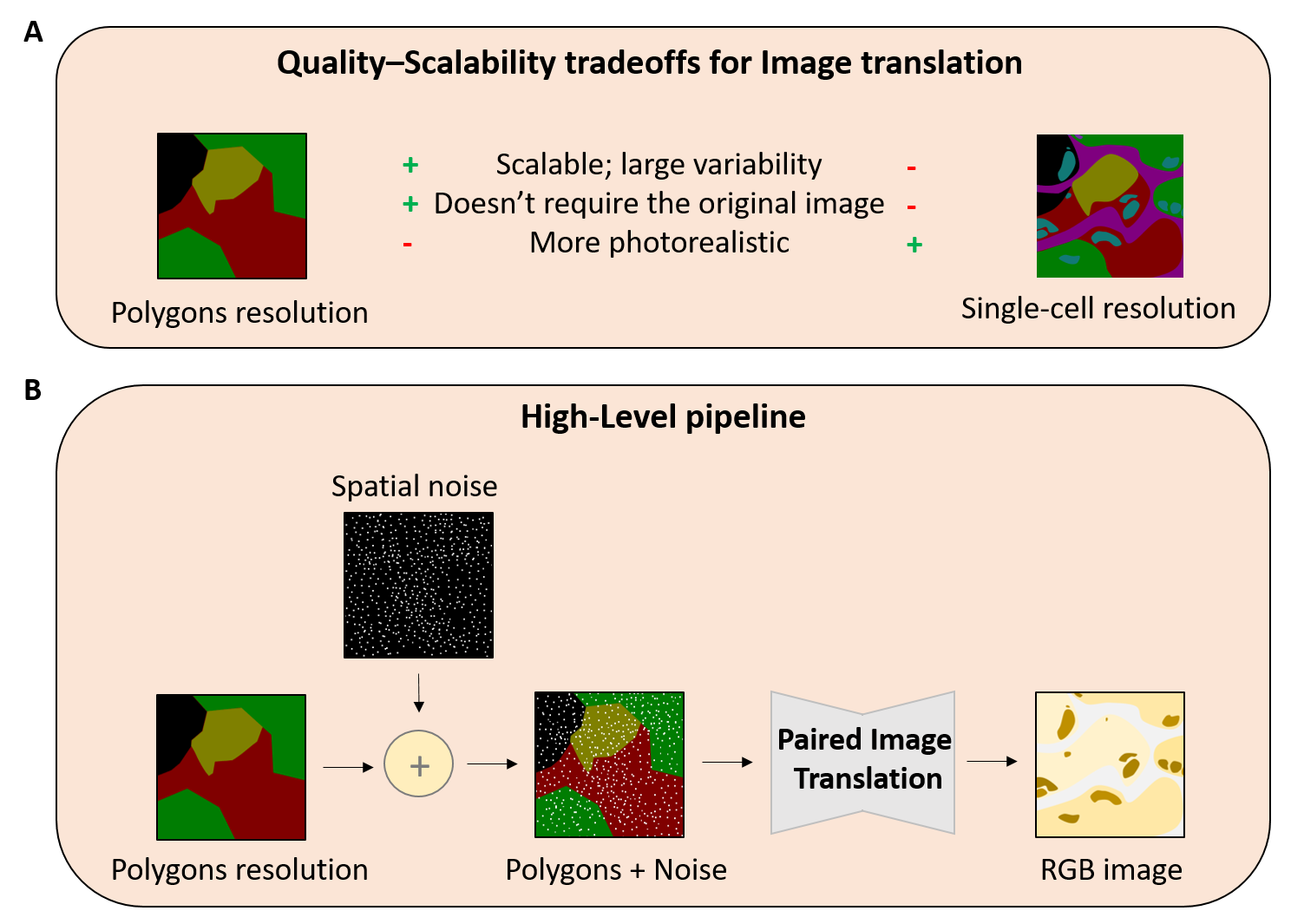}
	\caption{(A) In the case of Non-small cell lung carcinoma (NSCLC) the treatment choice is determined by the relative areas of three types of cell types that are annotated by pathologists manually: PD-L1 positive, PD-L1 negative, and Inflammation. Using these types of polygons maps as input for synthetic images allows control over the cellular features in mass. However, these smooth polygon regions pose a challenge for creating non-repetitive synthetic images. Adding detailed semantic masks can break the repetitive artifacts of the generated image. Yet, this fragmentation requires detailed knowledge of the actual cell distribution in the real images and therefore is not scalable. (B) Our pipeline uses an image translation methodology by generating a histology photorealistic image from a given semantic label mask. As an intermediate step, we add a random noise with different frequencies to the semantic mask, where the spatial noise frequency is a hyperparameter of the model. Next, we apply the semantic mask with the additional label to the image translation generator to produce the synthetic histology image.}
	\label{f:f2}
\end{figure}

In the case of tissues, both approaches have been harnessed with limited success. While the classic approach can result in photo-realistic images, the ability to control the distribution of objects within the images themselves (such as the location of the cell, blood vessels, etc.) is limited. When producing the masks that would be used as input, the resolution of the semantic labeling is critical (Fig.~\ref{f:f2}A). Typical masks of histology images contain only regional data (i.e. polygons that engulf regions of some cell types). The reason for that is that generating input masks that contain full single-cell information is challenging. Generating polygon input masks allows control over the cell types in the synthetic images, and allows scalability. However, polygon masks containing large smooth areas can result in repetitive artifacts and hinder the photorealism of the synthetic image.

In this work, we show that introducing random noise in particular frequencies into polygon-based masks can improve dramatically the quality of synthetic images, resulting in an image quality that is almost as good as providing the single-cell structure (Fig.~\ref{f:f2}B). We test our pipeline on Immunohistochemistry (IHC)-treated lung biopsies Lung cohort using a three-fold validation approach: Image similarity, Turing test, and AI improvements. Our work demonstrates how synthetic images can be easily created from masks that contain only regional data. These results pave the way for an automated diagnosis of Non-Small Cell Lung Cancer (NSCLC) and can be utilized for other conditions with similar challenges.

\section{RELATED WORK}
\subsection{Medical Image Synthesis}
Generative adversarial networks (GANs) \citep{goodfellow2020generative} aim to model the distribution of real images given the input noise via a minimax game between a generator, $G$, and a discriminator, $D$. Where the $G$ tries to generate synthetic images as close to the real images as possible whereas $D$ tries to distinguish them apart.
Conditional Generative adversarial networks (CGAN) \citep{mirza2014conditional} is a type of GAN that allows controlling the spatial distribution of the generated data by providing conditioning information.\\
Medical AI researchers have leveraged these networks, whose goal is to generate large and diverse datasets for training and evaluating deep networks \citep{shen2017deep}. Hence, it enables a wide variety of applications in medical imaging, and in digital pathology in particular, such as image generation (Breast cancer \citep{ma2018novel_relatedwork}, Glioblastoma \citep{janowczyk2016deep_relatedwork}, Colon cancer \citep{wang2020application_relatedwork}, CT scans from MRI \citep{peng2021knowledge_p2p_cyclegan_relatedwork}, Skin lesion \citep{bissoto2018skin}, Retinal fundi images \citep{guibas2017synthetic}), image adaptation \citep{chen2020unsupervised_cyclegan_relatedwork}, image enhancement \citep{zanjani2018stain_p2p_p2phd_cg_relatedwork}, and representation learning \citep{quiros2019pathologygan}. Yet, there are several challenges that need to be addressed in the generation of synthetic medical images. First, the images must have realistic texture, and second, they must be representative of a wide variety of tissue types and pathological conditions, such as the amount and location of cancer or immune cells.

\subsection{Image Translation}
The image-to-image translation is a set of tasks that translate the source domain of images to the target domain, either given input-output image training pairs, such as pix2pix \citep{p2p_paper}, pix2pixHD \citep{p2phd_paper} or without any explicit correspondence between the images in the two sets such as CycleGAN \citep{cyclegan_paper}. These deep learning models are typically using a variant of image-conditional GANs \citep{mirza2014conditional}. One of these kinds of tasks is to insert a semantic map and translate it to an image based on the additional information, such as class labels, passed together with the image to the network during the training phase. Thus generating images from labels in a wide variety of applications and domains in an effective manner.

\subsection{AI-assisted NSCLC diagnosis}
Non-small cell lung cancer (NSCLC)  is the most common type of lung cancer, accounting for $85\%$ of all lung cancer cases \citep{ettinger2008non}. While recent development in Immunotherapy has shown promising results in treating NSCLC \citep{dellon2010inter}.

One of the most common ways to assess cancer's stage and characteristics is based on tissue scans. Those are mostly being decoded by pathologists. According to their diagnosis, the treatment plan is determined. For example, some patients can be treated with immunotherapy methods. Those methods are tremendously helpful for some patients but can be harmful to others, and they are also very expensive. Hence, there is an urgent need for identifying responders and non-responders at an early stage \citep{adam2019p2}.
Pathologists usually use IHC-stained methods \citep{key2006immunohistochemistry} to decide whether this treatment is beneficial or not. IHC slides emphasized the expressions of Programmed Death-Ligand 1 (PD-L1), which is usually amplified by cancer cells. PD-L1 neutralizes white blood cells' activity, thereby causing the immune system to ignore the cancerous cells. Hence, the cancerous cells exude the PD-L1, which performs in two different expressions, such as NSCLC PD-L1 positive and NSCLC PD-L1 negative. The fraction of PD-L1 positive out of the total cancer PDL-L1 cells measured as the tumor proportion score (TPS) \citep{koelzer2019precision_43}. Its value divides the patients into 3 classes: ($0\%-1\%$, $1\%-50\%$, $50\%-100\%$). 

Pathologists estimate this value by looking at the WSI themselves \citep{troncone2017reproducibility_44, cooper2017intra_45}. This estimation can be reliable when the labeling is clear, but the middle regions' assessments (around TPS=$1\%$, TPS=$50\%$) are not good enough. 

AI methods can then come forward and supply easy and more robust assessments in a wide variety of applications such as PD-L1 image classification \citep{sha2019multi_pdl1ai1}, PD-L1 image segmentation 
\citep{cui2019deep_pdl1ai2}, TPS severity classification \citep{kapil2018deep_dpl1ai3}, and also other realizations in digital pathology \citep{larey2022,czyzewski2020}.

\section{MATERIALS AND METHODS}
\subsection{Study population and dataset}
22 whole slide images (WSIs) from 19 patients that were stained using anti-PD-L1 antibody clone 22C3 Dako using a Ventana immunostainer following a harmonization procedure. The slides were scanned using PANNORAMIC 250 Flash III (3DHISTECH) at 40X. All procedures performed in this study and involving human participants were in accordance with the ethical standards of the Rambam Medical center institutional research committee,  approval 0522-10-RMB, and with the 1964 Helsinki declaration and its later amendments or comparable ethical standard.

For our analysis, we cropped a small set of $512$ images out of $22$ WSIs of NSCLC tissue samples with a size of 512X1024 pixels for semantic labeling. These images were manually annotated by $4$ trained and experienced researchers and were validated by an expert pathologist. Each pixel was assigned to one of four classes: NSCLC with PD-L1 expression ($n = 1281$ polygons, $n = 39.5M$ pixels), defined by PD-L1 positive, NSCLC without PD-L1 expression ($n = 871$ polygons, $n = 20.6M$ pixels), defined by PD-L1 negative, inflammation ($n = 1209$ polygons, $n = 29.5M$ pixels), and Other ($n = 172$ polygons, $n = 134.6M$ pixels), defined by healthy tissue and air.

In terms of Lung Cell TPS distribution, this dataset is imbalanced. Most of the images fall into a bimodal distribution with a lot of images getting TPS of $0$ (about $26.4\%$ of all dataset), and a lot getting TPS of $1$ (about $36.8\%$ of all dataset). On the one hand, these images have mostly NSCLC PD-L1 negative, or mostly NSCLC PD-L1 positive, but on the other hand around the clinical decision thresholds ($TPS=0.01$, $TPS=0.5$) there is less abundance of images (about $4\%$ of all dataset). To avoid training bias, the images were manually split to build a non-biased training set ($n = 360$ images) and test set ($n = 152$ images). We used this dataset to model the generation of synthetic biopsy images.

\subsection{Semantic segmentation metrics}
To estimate the UNet++ \citep{Unet_pp_paper} segmentation performances, we used the following metrics, 
\begin{equation}
mIoU = \frac{1}{I \cdot C} \sum_{i}\sum_{c} \frac{TP_{i,c}}{TP_{i,c}+FP_{i,c}+FN_{i,c}} \label{eq_miou}
\end{equation}

\begin{equation}
wIoU = \frac{1}{I \cdot C \cdot S} \sum_{i}\sum_{c} s_c \cdot \frac{TP_{i,c}}{TP_{i,c}+FP_{i,c}+FN_{i,c}} \label{eq_wiou}
\end{equation}

\begin{equation}
wPrecision = \frac{1}{I \cdot C \cdot S} \sum_{i}\sum_{c} s_c \cdot \frac{TP_{i,c}}{TP_{i,c}+FP_{i,c}} \label{eq_wprecision}
\end{equation}

\begin{equation}
wRecall = \frac{1}{I \cdot C \cdot S} \sum_{i}\sum_{c} s_c \cdot \frac{TP_{i,c}}{TP_{i,c}+FN_{i,c}} \label{eq_wrecall}
\end{equation}

\begin{equation}
tObjective = \frac{1}{\frac{1}{C} \sum_{c} \frac{2 \cdot TP_{c}}{2 \cdot TP_{c}+FP_{c}+FN_{c}} -\frac{1}{4} \cdot \sum_{c}y_{o,c}\log(pr_{o,c})} \label{eq_trainobjective}
\end{equation}

where the $c$ index iterates over the different classes in the image, and the $i$ index iterates over the different images in the dataset. $p_{ct}$ denotes the number of pixels of class $c$ classified as class $t$. 
$s_c=\sum_{t}p_{ct}$ is the total number of pixels belonging to class c, and $S=\sum_c s_c$ denotes the number of all pixels. $pr_{o,c}$ denotes the predicted probability observation o is of class c, and $y$ is a binary indicator (0 or 1) if class label c is the correct classification for observation o.
$C$ is the total number of classes, and $I$ is the total number of images. 
$TP$, $TN$, $FP$, and $FN$ are classification elements that denote the true positive, true negative, false positive, and false negative of the areas of each image, respectively.

\subsection{Image quality assessment metric}
To estimate the image synthesis of pix2pixHD \citep{p2phd_paper} performances, we used the FID (Frechet inception distance) similarity metric, which is considered the gold-standard metric to date.\\
FID is a visual quality discriminator for comparing the quality of generated images to real images by comparing the feature vectors of the images in the feature space of a pre-trained Inception network \citep{Inceptionv3_paper}. It is based on the Fréchet distance between the two distributions of feature vectors, which measures how similar the two distributions are \citep{FID_metric}. A lower FID score indicates that the generated images are more similar to the real images.

\subsection{Training procedure}
The updated model was trained and optimized using Pytorch \citep{Pytorch_framework} framework on a single NVIDIA GeForce RTX A6000 GPU with 48GB GPU memory. During the training, different hyper-parameters were examined using Adam Solver \citep{Adam_paper} with beta1=0.5 and beta2=0.999, a minibatch of size 1, a learning rate of 2e-4, while we keep the same learning rate for the first 500 epochs, and linearly decay the rate to zero over the next 200 epochs. Weights were initialized from a Gaussian distribution with a mean of 0 and a unit standard deviation of 0.02. The optimization loss function contains two terms. First, for the discriminator which is an average discriminator prediction’s mean square error (MSE) between synthetic and real images. Second, for the generator that consists of the classic adversarial loss based on Binary cross-entropy (BCE), and two features-based matching losses that force the output synthetic image to seem like the specific real image and thus keep the conditional features of the images. While all the loss function elements were weighted with values of one.

\subsection{The pix2pixHD formulation} \label{p2phd_section}
In this work, we used pix2pixHD \citep{p2phd_paper}, which is a conditional GAN framework for image-to-image translation, to generate synthetic pathological images. The pix2pixHD is an extension of the pix2pix model \citep{p2p_paper}, and generates high-resolution images, and better visual quality. This network has novel multiscale generators and discriminators, which contribute towards the stabilization and optimization of the training of conditional GANs \citep{mirza2014conditional} on high-resolution images, and thus aims to achieve state-of-the-art results of fine geometry-image details and realistic textures.\\ 
Particularly, in generator $G$ architecture, we used only a single $G1$ that focus mainly on producing low-resolution images of size 512X1024 pixels based on global information, out of the decomposition of multiscale generators ($G1$, and $G2$). In Discriminator $D$ architecture, we used two multiscale discriminators ($D1$, and $D2$) with the same architecture, but works on different image scales, out of the decomposition of three discriminators ($D1$, $D2$, and $D3$). Hence, the $D$ aims to distinguish not only between a real and synthetic image in the entire image, but also in the fine details and the different textures. As a result, the $G$ is forced to study the true distribution of information on all scales, thus obtaining higher-quality images even in the smallest details.
Hence, the objective of the pix2pixHD model is expressed as:
\begin{equation}
\min_{G}((\max_{D_{1,..,K}}\sum_{k=1}^{K}\mathcal{L}_{GAN}(G,D_k))+\lambda \cdot \sum_{k=1}^{K}\mathcal{L}_{FM}(G,D_k))) \label{eq_p2phd}
\end{equation}

where $\lambda$ is a regularization parameter, and $K$ is the number of discriminators that have an identical deep network structure but operate at different image scales. In our study, we used $K=2$, which refers to the discriminators as $D1$, and $D2$. $\mathcal{L}_{GAN}(G,D)$ is conditional GAN loss, $\mathcal{L}_{FM}(G,D)$ is a feature matching loss, both are described in (\ref{eq_gan}) and (\ref{eq_fm}), respectively.

\begin{equation}
\min_{G}\max_{D}\mathbb{E}_{s,x}[\log{D(s,x)}] +  \mathbb{E}_{s}[\log{(1 - D(s,G(s)))}] \label{eq_gan}
\end{equation}

where $G$ is a generator and $D$ is a discriminator. $s$ represents the semantic label map, $x$ is the real image, and $G(s)$ is the generated image given the prior $s$. In the first term, the expectation $\mathbb{E}_{s,x}$ is over both the real pairs of semantic priors and images and in the second term $\mathbb{E}_{s}$, is over the semantic priors alone.

\begin{equation}
\mathbb{E}_{s,x}\sum_{i=1}^{T}\frac{1}{N_{i}}[|| D^{i}(s,x)-D^{i}(s,G(s)) ||_1] \label{eq_fm}
\end{equation}

where the $i$th-layer feature extractor of discriminator $D$ as $D^{i}$.
$T$ is the total number of layers and $N_{i}$ denotes the number of elements in each layer.

\subsection{Semantic labeling resolutions for Image Synthesis}
In this work, we compared three different resolutions of the semantic labeling for the generation of synthetic images of IHC-treated lung biopsies. All the resolutions are based on the pix2pixHD model described in subsection \ref{p2phd_section}. The only difference between them is the input mask that contains the conditions to generate the synthetic images. The three approaches considered in this work are the following:
\begin{itemize}
    \item Polygons' mask is a typical mask of histology images containing only regional data.
    \item Polygons + Noise mask is a noisy mask of histology images containing regional data with random Gaussian noise.
    \item Polygons + Air + Cells mask is a single cell mask of histology images containing air (non-tissue regions), single cells, and NSCLC feature regions data.
\end{itemize}

Where Polygons mask creation only needs the manually annotated NSCLC feature classes (PD-L1 positive, PD-L1 positive, Inflammation, Other) to obtain the corresponding input mask. Polygons + Noise masks are superpositions of the Polygons' mask with a random Gaussian noise, which is easy to generate automatically by a primary array programming library. Polygons + Air + Cells masks need in addition to Polygons annotations, the original RGB image information to extract the air and cells that forces the tissue mask structure to be similar to the original image. To extract air and cells from tissue images, we used classical computer vision methods to convert the images to grayscale and apply thresholds to extract air and cell pixels to distinguish between air and intracellular pixels.

\subsection{Pipeline Architecture}
Our pipeline for generating synthetic biopsy Images builds upon the pix2pixHD model. While \citep{p2phd_paper} uses instance-wise features in addition to labels as an input to image generation network $G$, we use the Gaussian random noise in addition to labels. Since NSCLC PD-L1 semantic label maps have a small number of classes and contain typically large and uniform polygons, random noise addition enables to challenge of the image generation process by avoiding repetitive texture effects, thereby achieving better image quality.

\begin{figure}[thpb]
\centering
	\includegraphics[scale=0.7]{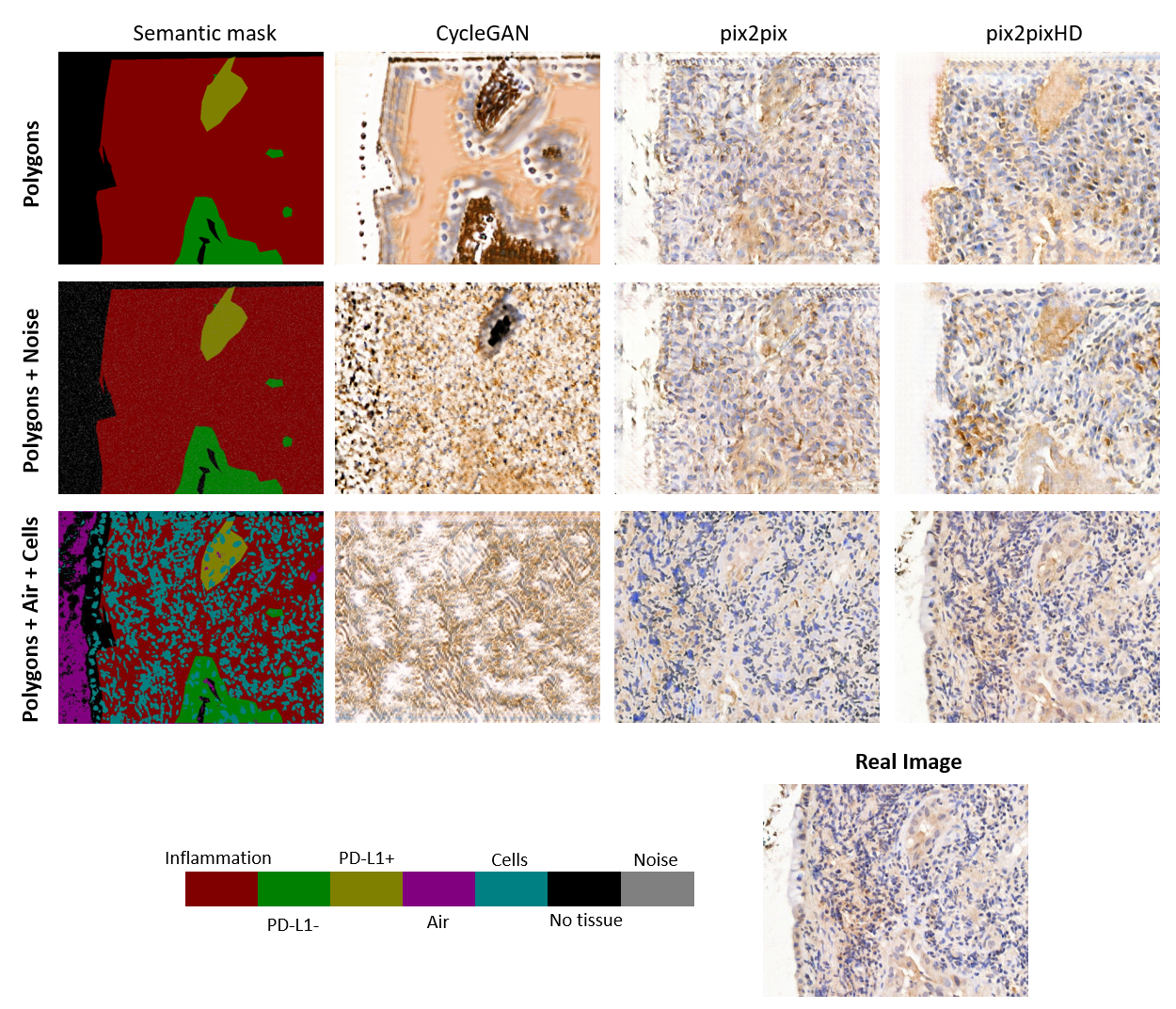}
	\caption{IHC-stained of NSCLC synthetic images from semantic layouts of 512X1024 pixels. Visual comparison of three types of conditional Image-to-Image translation approaches, which were used for producing synthetic images, show that pix2pixHD outperforms CycleGAN and pix2pix models. On the other hand, introducing random noise as an additional label which is spread in the entire image spatially eliminates blur and repetitive artifacts and improves the synthetic tissue fine details compared to the base polygons’ image resolution, similar to the level of adding single-cell resolution labeling.}
	\label{f:f3}
\end{figure}

\section{RESULTS}
\subsection{Comparison of different approaches}
To test the effect of adding noise to the semantic masks, we compared several image translation approaches for visual inspection of the generated histology synthetic images. The approaches included CycleGAN \citep{cyclegan_paper}, pix2pix \citep{p2p_paper}, and pix2pixHD \citep{p2phd_paper} models. 
We compared three types of semantic masks: 1) with Polygons, 2) Polygons + Noise, and 3) Polygons + Air + Cells.  Generated tissue image based on Polygons contains blur and repetitive artifacts due to the large smooth areas and can be explained by pix2pixHD fractionally-stride convolution architecture. When using Polygons + Air + Cells masks, masks that carry a lot of prior information, images  have high similarity to the original images, therefore are more photorealistic, but not scalable for improving algorithms and existing AI models. 

Hence, in the context of the quality-scalability trade-off, Polygons + Noise masks help to provide not only high-quality images with tissue fine details similar to the level of single-cell resolution labeling (Polygons + Air + Cells), but also add more control over the image. Therefore, we can conclude that Polygons + Noise masks allow for generating an easily more diverse set of high-quality images, and avoiding the time-consuming of manual image annotation.

\begin{figure}[thpb]
\centering
	\includegraphics[scale=0.45]{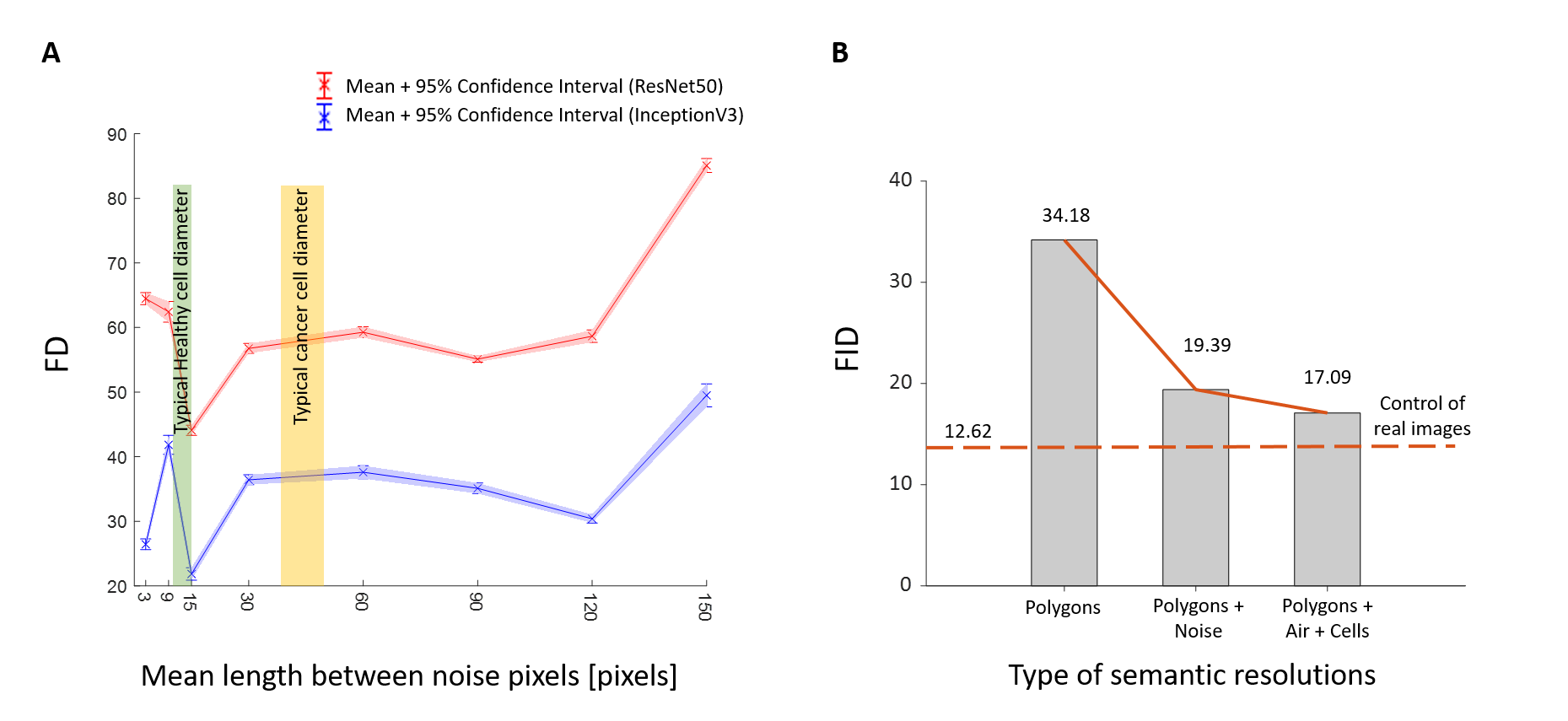}
	\caption{The effect of noise on image similarity (A) Eights different pix2pixHD models were trained with different spatial noise  to determine the optimal noise frequency. Applying a visual quality discriminator, such as Fréchet Distance (FD) based on ResNet50 and InceptionV3 deep architectures, on the same set of (n= 152) test real images, shows that a mean 15 pixels between noise pixels yield the highest similarity comparison of the synthetic images to real images. The horizontal bars illustrate the typical size of healthy (green) and cancer cells (yellow). (B) Adding random spatial noise with the optimal frequency of 15 compared to the ideal case of single-cell resolution (i.e. Polygons + Air + Cells), while the reference line represents the best similarity score of 12.62 can be achieved on the same test set based on control real images. The marginal improvement of adding noise is almost the same as in the case of adding real single-cell features.}
	\label{f:f4}
\end{figure}

\subsection{Random noise frequency optimization}
To test the effect of noise frequency, eight different pix2pixHD models were trained with  different mean distances between noise pixels. To test the similarity of the synthetic images we used  InceptionV3 \citep{Inceptionv3_paper} and ResNet50 \citep{Resnet50_paper}, and evaluated the performance of $n = 152$ synthetic images using a visual quality discriminator, based on Frechet Distance (FD) \citep{FD_metric}. Analyzing the results, we can observe that a mean length of $15$ pixels between two noise pixels yields the highest similarity comparison of the synthetic images to real images on both architectures (shown in Fig.~\ref{f:f4}A).

A mean length of $15$ pixels is within the range of the characteristic frequency of the healthy cell to the cancer cell in the realization of NSCLC. Fig.~\ref{f:f4}B presents a similarity comparison of generated synthetic images from different semantic labeling resolutions (in terms of FID \citep{FID_metric}). It can be shown that adding random noise to polygon-based masks is closer to the result based on a single-cell structure, than the synthetic images generated by polygon-based masks by a factor of $1.76$.

\subsection{Algorithmic Improvement and Turing Test}
\begin{figure}[thpb]
\centering
	\includegraphics[scale=0.4]{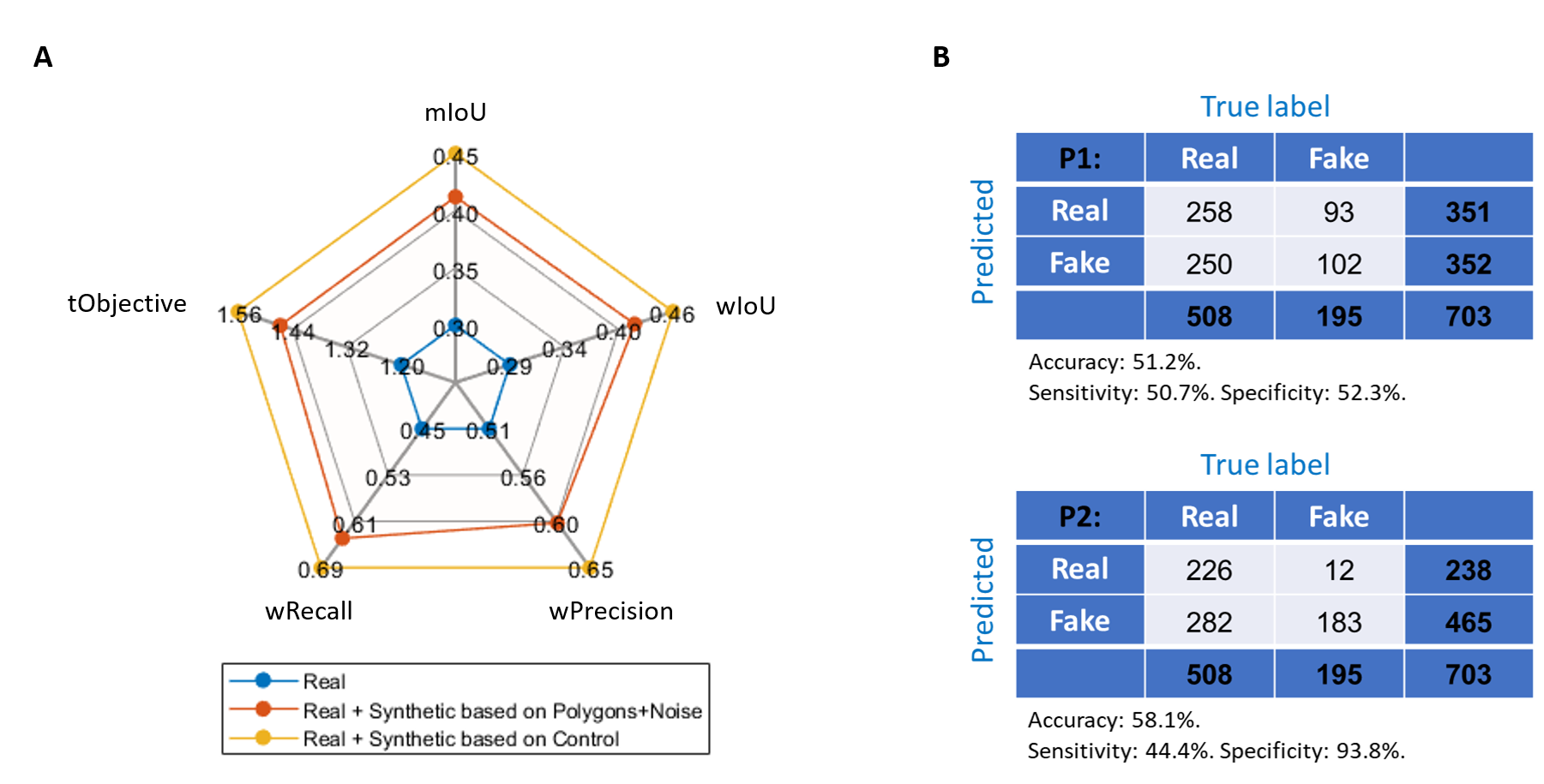}
	\caption{AI validation and Turing test of the NSCLC synthetic images. (A) To test the effect of synthetic images on AI performances, state-of-the-art architecture (UNet++) was trained on 100 real images and validated on a different real 100 ones. Adding 152 synthetic images based on polygon + noise masks to the real training set (red) improves the semantic segmentation accuracy by over 17\% compared to baseline test results (blue), and a slightly lower than adding the same amount of control / different set of real images (yellow). (B) Two trained, experienced researchers were presented with both real and synthetic images. This figure presents the Turing test results. P1, Expert 1 \#1; P2, Expert 2 \#2, IoU, intersection over union; m, mean; w, weighted; t, train.}
	\label{f:f5}
\end{figure}

State-of-the-art segmentation architecture, UNet++ \citep{ Unet_pp_paper}, was trained on $100$ real images followed by the \citep{pecnet_paper} hyperparameters, as a baseline model to distinguish between the four types of tissue cells, such as NSCLC PD-L1 positive, NSCLC PD-L1 negative, Inflammation cells, and other cells. We show in Fig.~\ref{f:f5}A, that adding ($n = 152$) synthetic images to the training set, improved the unseen test sample ($n = 100$) performance with respect to all segmentation metrics, described by (\ref{eq_miou})-(\ref{eq_trainobjective}). For instance, the network, fed by synthetic images, achieves better segmentation by a factor of $36.8\%$ and $17.8\%$ than the baseline model, in terms of mIoU and wPrecision, respectively.\\

In addition, we performed a Turing test where trained and experienced researchers were presented with both real and synthetic images, and Fig.~\ref{f:f5}B presents through the confusion matrices that our pipeline produces synthetic images reliable enough and with high quality similar to real images. This performance analysis is critical to the ability of the model network to be more robust and provide a reliable TPS for patient severity diagnosis.

\subsection{Unbiasing of IHC-treated lung datasets}
\begin{figure}[thpb]
\centering
	\includegraphics[scale=0.55]{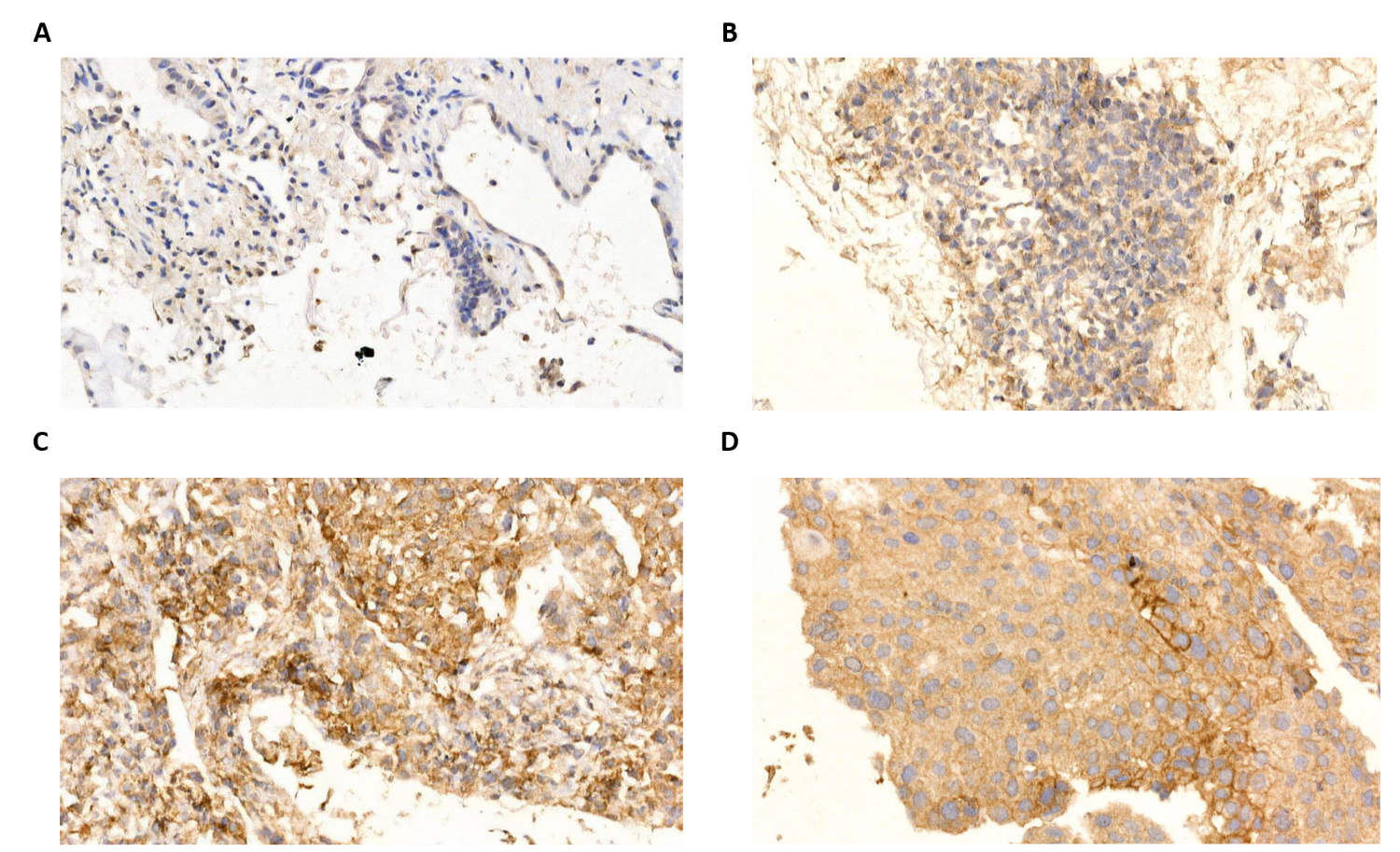}
	\caption{Controlling tumor proportion score (TPS) of synthetic images of immunohistochemistry-treated lung biopsies. (A) NSCLC healthy tissue image (TPS of 0). (B) NSCLC inflammatory tissue image (TPS of 0). (C) NSCLC with tumor markers (TPS of 1\%-50\%). (D) NSCLC with tumor markers (TPS of 100\%).}
	\label{f:f6}
\end{figure}

One of the main challenges in harnessing AI for TPS predictions is the bias within the data. Fig.~\ref{f:f6} demonstrates our ability to create synthetic images of any desired TPS.

\section{CONCLUSION}
Artificial Intelligence has the potential to revolutionize digital pathology by automating certain tasks and increasing the speed and accuracy of diagnosis. The use of synthetic data is becoming increasingly important for the development and training of AI algorithms in digital pathology. This type of data can be used to train AI algorithms in a controlled and efficient manner, without the need for real patient data. This is particularly beneficial in the field of digital pathology, where access to high-quality, annotated data can be limited. With synthetic data, researchers can generate large amounts of data that can be used to train AI algorithms and evaluate their performance. Additionally, synthetic data can be used to test the robustness of AI algorithms and identify potential issues before they are deployed in a clinical setting.

One of the main limitations of achieving debasing of histological datasets using synthetic images is the need ability to control the feature distribution in a precise manner. Conditional GANs and in particular paired GANs can provide such control but unlike other domains, such as autonomous vehicles or face recognition, in the case of tissues generating the 'scene' is a challenge by itself. One way to control the features of the synthetic images is to define regions of cell types by using polygon-based semantic masks as inputs. However, this approach can lead to inherent  artifacts that are the result of generating a pattern with a small scale (that is the single-cell scale) with a smooth area input (the polygon that marks the area of the cells).

In this work, we show that introducing single-pixel random noise with a mean distance that is within the typical scale of cells, can remove these artifacts. We demonstrate that adding a random noise is almost equivalent to adding the actual single-cell information itself. Therefore, our approach can use polygon semantic masks and noise to create images with any desired tumor-proportion score. Moroever, these images are not only similar to the real ones in terms of similarity metrics but also to human experts, and can be used to improve AI performance. 

Our results demonstrated the ability to overcome the problem of biased datasets in such as the frequency of rare disease cases, and cases that are at the critical thresholds of clinical decisions. In addition, it facilitates digital pathology AI development for histopathology diagnosis by improving AI models' performance, and robustness and understanding their failure cases. 

\section*{ACKNOWLEDGMENT}
The authors would like to thank Tanya Wasserman, Tal Ben-Yaakov, Yair Davidson, and Yael Abuhatsera for their technical support and valuable discussions.

\bibliographystyle{plainnat}

\end{document}